\documentclass[11pt,a4paper]{article}

\usepackage{jheppub}
\usepackage{amsmath}
\usepackage{graphicx}
\usepackage{slashed}
\usepackage[utf8]{inputenc}
\usepackage{caption}
\usepackage{subcaption}
\usepackage{tikz}

\newcommand{\ep}{\epsilon}

\newcommand \vev [1] {\langle{#1}\rangle}
\newcommand \ket [1] {|{#1}\rangle}
\newcommand \bra [1] {\langle {#1}|}
\newcommand{\cN}{{\cal N}}

\def\eps{\epsilon}

\allowdisplaybreaks[3]

\makeatletter
\renewcommand\@fpheader{} 
\renewcommand\@journal{}
\makeatother

\title{Analytic result for the nonplanar hexa-box integrals}

\preprint{{ZU-TH 35/18, MPP-2018-220,  MITP/18-084, IPPP/18/75}}

\author[a]{D.~Chicherin,}
\author[b]{T.~Gehrmann,}
\author[a,c]{J.M.~Henn,}
\author[d]{N.A.~Lo~Presti,}
\author[a]{V.~Mitev,}
\author[a]{P.~Wasser}

  \affiliation[a]{
  PRISMA Cluster of Excellence, Institute of Physics,
  Johannes Gutenberg University,\\
  D-55099~Mainz, Germany}
  
\affiliation[b]{
  Physik-Institut, 
  Universit\"at Z\"urich, Wintherturerstrasse 190,
  CH-8057~Z\"urich, Switzerland}

  \affiliation[c]{
  MPI f\"ur Physik, 
  Werner-Heisenberg-Institut,\\
  M\"unchen, Germany}

  \affiliation[d]{
  Institute for Particle Physics Phenomenology, Durham University,
  Durham DH1 3LE, UK}  

\emailAdd{chicherin@uni-mainz.de}
\emailAdd{thomas.gehrmann@uzh.ch}
\emailAdd{henn@uni-mainz.de}
\emailAdd{nicola.a.lo-presti@durham.ac.uk}
\emailAdd{vmitev@uni-mainz.de}
\emailAdd{wasserp@uni-mainz.de}


\abstract{
In this paper, we analytically compute all master integrals 
for one of the two non-planar integral families for five-particle massless scattering at two loops.
We first derive an integral basis of 73 integrals with constant leading singularities.  
We then construct the system of differential equations satisfied by them, and find that it is in canonical form.
The solution space is in agreement with a recent conjecture for the non-planar pentagon alphabet. We fix the boundary constants of the differential equations by 
exploiting constraints from the absence of unphysical singularities.
The solution of the differential equations in the Euclidean region is expressed in terms of iterated integrals.
We cross-check the latter against previously known results in the literature, as well as with independent Mellin-Barnes calculations.
}

\begin{document}
\unitlength1cm
\maketitle

\newpage

\section{Introduction}
\label{sec: intro}

Scattering amplitudes for multi-particle processes start 
to play an increasingly important role in future collider physics analyses, as processes 
at higher multiplicity are being probed more and more accurately. 
Recently, rapid progress has been achieved for five-particle processes at next-to-next-to-leading order. This concerns several areas, such as the efficient computation of loop integrands \cite{Badger:2013gxa,Ita:2015tya,Abreu:2017idw,Badger:2017jhb}, 
the analytic computation of the Feynman integrals \cite{Gehrmann:2015bfy,Papadopoulos:2015jft,Gehrmann:2018yef} as well as advances in integral reduction techniques 
\cite{Gluza:2010ws,Schabinger:2011dz,vonManteuffel:2014ixa,Larsen:2015ped,Peraro:2016wsq,Kosower:2018obg,Boehm:2017wjc,Boehm:2018fpv,Chawdhry:2018awn}. 
Most recently, two independent numerical determinations of all 
planar five-gluon scattering amplitudes \cite{Badger:2017jhb,Abreu:2017hqn} have been achieved. 

Non-planar corrections are unfortunately considerably more difficult to handle, due to a variety of reasons. Owing 
to the richer cut-structure of non-planar amplitudes, they can contain a larger number of rational factors in the 
external invariants, leading to more complicated algebraic expressions, both in the 
integrand reduction and in the determination of the integrals. This 
 article addresses the second challenge, specifically at the level of the computations of non-planar Feynman integrals.
The first steps in this direction were taken in \cite{Chicherin:2017dob}, where three of the present authors
conjectured the function space describing the Feynman integrals, and proposed a bootstrap method
for determining the functions.
Furthermore, individual integrals were computed in ref.~\cite{Chicherin:2018ubl}, using a method based on conformal symmetry.

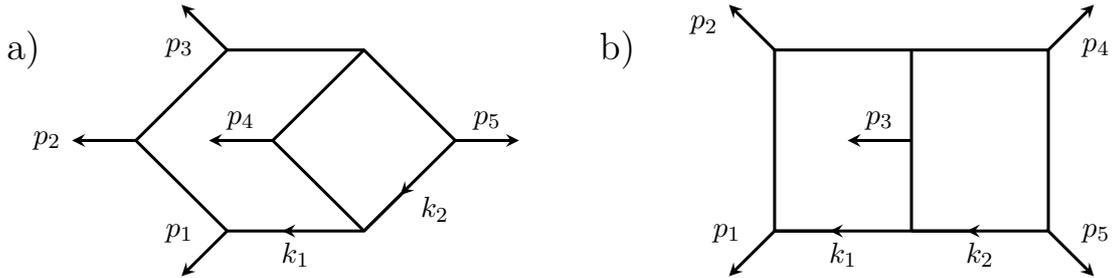
\begin{figure}
\centering
\begin{tikzpicture}[scale=.6, line width=0.4mm, >=stealth]
 \begin{scope}    [shift={(-2,0)}]

 \node[text width=1cm] at (-3,4) {{\Large a)}};

 \draw[-<] (1,0) -- (2.5,0) node[pos=1.0, below]{$k_1$};
 \draw[-] (2.3,0) -- (4,0) -- (6,2) -- (4,4) -- (1,4) -- (-1,2) -- (1,0);
 \draw[-<] (4,0) -- (5,1) node[pos=1.0, below right]{$k_2$};
 \draw[-] (4,0    ) -- (2,2) -- (4,4);
 \draw[->] (1,0) -- (0,-1) node[pos=.5,above left]{$p_1$};
 \draw[->] (-1,2) -- (-2.4,2) node[pos=1.,left]{$p_2$};
 \draw[->] (1,4) -- (0,5) node[pos=.5,below left]{$p_3$};
 \draw[->] (2,2) -- (0.6,2) node[pos=.5,above]{$p_4$};
 \draw[->] (6,2) -- (7.4,2) node[pos=.5,above]{$p_5$};
 \end{scope}
 \begin{scope}[shift={(11,0)}]
   \node[text width=1cm] at (-3,4) {{\Large b)}};
 \draw[-] (0,0) -- (6,0) -- (6,4) -- (0,4) -- (0,0);
 \draw[-] (3,0) -- (3,4);
 \draw[-<] (0,0) -- (1.5,0) node[pos=1.,below]{$k_1$};
 \draw[-<] (3,0) -- (4.5,0) node[pos=1.,below]{$k_2$};
 \draw[->] (0,0) -- (-1,-1) node[pos=.5,above left]{$p_1$};
 \draw[->] (0,4) -- (-1,5) node[pos=1.,below left]{$p_2$};
 \draw[->] (6,4) -- (7,5) node[pos=.5,below right]{$p_4$};
 \draw[->] (6,0) -- (7,-1) node[pos=.5,above right]{$p_5$};
 \draw[->] (3,2) -- (1.6,2) node[pos=.5,above]{$p_3$};
 \end{scope}
\end{tikzpicture}
\caption{On the left, with label a), we depict the hexa-box integral family and on the right, with label b), the double pentagon integral family.}
\label{fig:boxfamily}
\end{figure}

There are two non-planar integral families for five particles at two loops, namely the hexa-box integral family a) and the double pentagon integral family b), shown in Fig.~\ref{fig:boxfamily}.
In this paper, we analytically compute all master integrals for this first one, namely the hexa-box integral family a).

We begin by deriving a basis of integrals with constant leading singularities, also known as d-log integrals \cite{ArkaniHamed:2010gh,Henn:2013pwa,Bern:2015ple}.
This is done by adapting the algorithm described in \cite{WasserMSc} to the five-particle kinematics.
We then use integral reduction programs to find a basis of $73$ d-log integrals.  

We follow this up by computing the differential equations for the basis integrals, 
and find that they obey the canonical form of \cite{Henn:2013pwa},
as expected by the conjecture made therein. We find that the differential 
equations can be expressed in terms of the non-planar
pentagon alphabet of reference \cite{Chicherin:2017dob}.

Having obtained a system of first-order differential equations, the solution is fully specified by providing a 
complete set of boundary constants.
We do so by deriving constraints from the absence of unphysical singularities. 
In this way, we obtain analytical constraints for all boundary constants 
(up to an overall normalization, which is fixed by a trivial calculation).
The constraints are written in terms of Goncharov polylogarithms. We evaluate the latter to 
high numerical precision, and give the
solutions with 100-digit accuracy.
This fully determines the solution of the differential equations,  which can be expressed in terms of iterated integrals. These are straightforward to evaluate in the Euclidean region, as documented in detail for example in~\cite{Gehrmann:2018yef}. To obtain the results in the Minkowskian region 
requires either an analytic continuation of the results, or an independent determination of the boundary conditions in each Minkowskian scattering channel.

We validate our solution by comparing it to previously known results in the literature, for subtopologies that are 
planar or that correspond to four-point functions,
as well as against an independent Mellin-Barnes calculation described in appendix~\ref{sec:mellinbarnes}.

The paper is organized as follows. We begin in section~\ref{sec:kinematics} by describing our notation and
the kinematics of the problem. We also discuss the integral reduction to master integrals, and the differential equation satisfied by the latter. 
Then, in section~\ref{sec:dlog}, we explain the determination of the d-log basis.
Section \ref{sec:de} is dedicated to the canonical differential equations and their analytic solution. The appendix \ref{sec:mellinbarnes} 
discusses checks performed on the results. Finally, we conclude in section \ref{sec:conclusion}.

\section{Setup}
\label{sec:kinematics}

\subsection{Kinematics and notation}

We denote the momenta of the on-shell particles in pentagon kinematics
 by $p^\mu_{i}$, $i=1,\ldots 5$, with $p_{i}^2=0$.
Momentum conservation reads $\sum_{i=1}^{5} p^{\mu}_{i}=0$.
We introduce the following five independent Mandelstam variables,
\begin{align}
\label{eq: definition vi}
v_{1} = \frac{2 p_{1} \cdot p_{2}}{\mu^2} \,,\quad 
v_{2} = \frac{2 p_{2} \cdot p_{3}}{\mu^2} \,,\quad
v_{3} = \frac{2 p_{3} \cdot p_{4}}{\mu^2} \,,\quad
v_{4} = \frac{2 p_{4} \cdot p_{5}}{\mu^2} \,,\quad
v_{5} = \frac{2 p_{5} \cdot p_{1}}{\mu^2} \,.
\end{align}
Here $\mu$ is an arbitrary scale, e.g.\ the scale appearing in the dimensional regularization, making the 
$v_i$ dimensionless.
In the following, we will set $\mu^2=1$~GeV without loss of generality, 
as the dependence on $\mu$ can always be restored by dimensional analysis.
%
We parametrize the integrals of the integral family shown in Fig.~\ref{fig:boxfamily}a) using the following notation
\begin{align}
\label{eq:definition_of_F}
 F_{a_1,...,a_{11}}\, =\,&  \int 
 \frac{d^Dk_1 d^Dk_2}{(i \pi^{D/2})^2} \frac{[(k_2-p_1)^2]^{-a_9}[(k_2-p_1-p_2)^2]^{-a_{10}}}{[k_1^2]^{a_1}[(k_1-p_1)^2]^{a_2}[(k_1-p_1-p_2)^2]^{a_3}[k_1-p_1-p_2-p_3)^2]^{a_4}} \notag \\
 &\times \frac{[(k_2-p_1-p_2-p_3)^2]^{-a_{11}}}{[k_2^2]^{a_5} [(k_2-p_1-p_2-p_3-p_4)^2]^{a_6}[(k_1-k_2)^2]^{a_7}[(k_1-k_2+p_4)^2]^{a_8}}
\end{align}
for the individual integrals. In the above, $a_1, \ldots, a_8 \geq  0$ are propagators and
$a_9$, $a_{10}$ and $a_{11}\leq 0$ numerator factors.

To perform the integral reduction \cite{Laporta:2001dd} for this hexa-box family, we use the program 
Reduze2 \cite{vonManteuffel:2012np}, which yields a basis of 73 master integrals. Aiming for 
the differential equations for the hexa-box family, we need to go beyond the reduction of 
integrals with unit powers on all propagators 
(which was accomplished previously,~\cite{Boehm:2018fpv,Chawdhry:2018awn}), 
which are sufficient for scattering amplitudes, and 
include the reduction of integrals with single squared propagators. 
To limit the size of intermediate algebraic expressions in this reduction, we perform independent reductions 
on spanning cuts, i.e.\ by projecting onto subspaces of integrals that are required to contain a specific 
combination of propagators. The hexa-box family has in total 11 spanning cuts (single-scale 
three-point or
four-point subtopologies that each  do not contain any further subtopologies). A sufficient 
practical mitigation of the 
 complexity of the integral reduction  
can be achieved by combining them into four spanning topologies, identified
 by two-particle cuts, i.e.\ requiring the non-vanishing of 
$(a_5,a_7)$, $(a_5,a_8)$, $(a_6,a_7)$ or $(a_6,a_8)$. The full integral reduction is then assembled by adding 
the  cut-reductions, with individual integrals weighted by appropriate inverse multiplicity factors, which correct for 
their appearance in more than one cut-reduction tree. 

\begin{figure}[t]
             \begin{center}       
              \includegraphics[width=13cm]{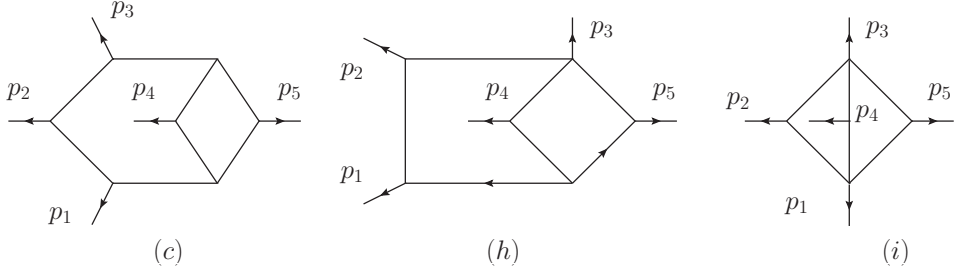}
              \caption{Non-planar integral sectors with genuine five-particle kinematics.
              The labelling follows that of \cite{Bern:2015ple}.}
              \label{Fig:non-planar integrals}
            \end{center}
\end{figure}

The master integrals in the hexa-box family 
can be classified as follows. There are 54 planar integrals, 9 are non-planar with up to four external legs 
(four-point functions with one off-shell leg, which were computed in~\cite{Gehrmann:2000zt, Gehrmann:2001ck, Gehrmann:2002zr} in terms of generalized harmonic polylogarithms \cite{Remiddi:1999ew, Goncharov:2001iea,Gehrmann:2001pz, Gehrmann:2001jv, Vollinga:2004sn}) and 10 that are non-planar with five external legs. 
The latter type of genuine non-planar five-point 
integrals in the hexa-box integral family are depicted in Fig.~\ref{Fig:non-planar integrals}. 
The second one, (h), can also be flipped upside down, and hence there are four such integral sectors. Together they have $3+3+3+1=10$ master integrals. The reduction selects a basis of master integrals in 
each topology according to lexicographic ordering, typically containing the scalar integral and integrals with simple 
numerator factors. 
Differential equations for the hexa-box integrals 
in an  alternative basis in terms of pure integrals (containing higher propagator powers) 
were derived most recently in \cite{Abreu:2018rcw}. 

We will be interested in a different basis, in which the integrals have a d-log form. Such d-log integrals have properties that significantly simplify their computation. In particular, in the $\epsilon$ expansion all such integrals evaluate to multiple polylogarithms of homogeneous weight.  Determining this basis
is the subject of section~\ref{sec:dlog}. We note already here that this basis choice can be done algorithmically \cite{WasserMSc} by analyzing just the loop {\it integrand}. 

\subsection{The alphabet} 
The 73 integrals that we shall compute can be expressed through iterated integrals of the type $\int d\log W_{i_1}\cdots \int d\log W_{i_L}$, where the algebraic functions $W_i$ of the kinematic variables are called letters.  The ensemble of the letters $\{W_i\}$ is called the alphabet of the 
problem under consideration. We recall the notation of \cite{Chicherin:2017dob},
where the $31$ letters of the non-planar pentagon alphabet were introduced. They fall into six classes of five letters $W_{1+i}$, $W_{6+i}$, $W_{11+i}$, $W_{16+i}$, $W_{21+i}$, $W_{26+i}$, with $i=0\ldots 4$, that are mapped into each other by cyclic permutations together with one lonely letter $W_{31}$. Explicitly, the first twenty letters are
\begin{align}
\label{fullpentagonalphabet}
&W_{1} = \; v_{1} \,, &
&W_{6} = \; v_{3} + v_{4} \,, &
& W_{11} = \; v_{1} - v_{4} \,, &  
&W_{16} = \; v_{1}+ v_{2}-v_4 \,, & \nonumber\\
&W_{2} = \; v_{2} \,, &
&W_{7} = \; v_{4} + v_{5} \,, &
& W_{12} = \; v_{2} - v_{5} \,, &  
&W_{17} = \;  v_{2}+ v_{3}-v_5 \,, & \nonumber\\
&W_{3} = \; v_{3} \,, &
&W_{8} = \; v_{5} + v_{1} \,, &
& W_{13} = \; v_{3} - v_{1} \,, &  
&W_{18} = \; v_{3}+ v_{4}-v_1 \,, & \\
&W_{4} = \; v_{4} \,, &
&W_{9} = \; v_{1} + v_{2} \,, &
& W_{14} = \; v_{4} - v_{2} \,, &  
&W_{19} = \; v_{4}+ v_{5}-v_2 \,, & \nonumber\\
&W_{5} = \; v_{5} \,, &
&W_{10} = \; v_{2} + v_{3} \,, &
& W_{15} = \; v_{5} - v_{3} \,, &  
&W_{20} = \; v_{5}+ v_{1}-v_3 \,, & \nonumber
\end{align}
while the next ten are
\begin{align}
&W_{21} = \; v_3 + v_4 - v_1 - v_2 \,, &  
&W_{26} = \; \frac{v_1 v_2-v_2 v_3+v_3 v_4-v_1 v_5-v_4 v_5-\sqrt{\Delta}}{v_1 v_2-v_2 v_3+v_3 v_4-v_1 v_5-v_4 v_5+\sqrt{\Delta}} \,, & \nonumber\\
&W_{22} = \; v_4 + v_5 - v_2 - v_3 \,, &  
&W_{27} = \; \frac{-v_1 v_2+v_2 v_3-v_3 v_4-v_1 v_5+v_4 v_5-\sqrt{\Delta}}{-v_1 v_2+v_2 v_3-v_3 v_4-v_1 v_5+v_4 v_5+\sqrt{\Delta}} \,, & \nonumber\\
&W_{23} = \; v_5 + v_1 - v_3 - v_4 \,, &  
&W_{28} = \; \frac{-v_1 v_2-v_2 v_3+v_3 v_4+v_1 v_5-v_4 v_5-\sqrt{\Delta}}{-v_1 v_2-v_2 v_3+v_3 v_4+v_1 v_5-v_4 v_5+\sqrt{\Delta}}\,, & \\
&W_{24} = \; v_1 + v_2 - v_4 - v_5 \,, &  
&W_{29} = \; \frac{v_1 v_2-v_2 v_3-v_3 v_4-v_1 v_5+v_4 v_5-\sqrt{\Delta}}{v_1 v_2-v_2 v_3-v_3 v_4-v_1 v_5+v_4 v_5+\sqrt{\Delta}} \,, & \nonumber\\
&W_{25} = \; v_2 + v_3 - v_5 - v_1 \,, &  
&W_{30} = \; \frac{-v_1 v_2+v_2 v_3-v_3 v_4+v_1 v_5-v_4 v_5-\sqrt{\Delta}}{-v_1 v_2+v_2 v_3-v_3 v_4+v_1 v_5-v_4 v_5+\sqrt{\Delta}}\,, & \nonumber
\end{align}
and the last one is 
\begin{align}
W_{31}=\sqrt{\Delta}\,.
\end{align}
Here, $\Delta$ is the Gram determinant that can be written explicitly as 
\begin{equation}
\label{eq: Gram Determinant}
\begin{split}
\Delta\,=\,&v_1^2 (v_2 - v_5)^2 + (v_2 v_3 + v_4 (-v_3 + v_5))^2\\& + 
 2 v_1 (-v_2^2 v_3 + v_4 (v_3 - v_5) v_5 + 
    v_2 (v_3 v_4 + (v_3 + v_4) v_5))
    \end{split}
\end{equation}
Note that the letters $W_{i}$, with $i=26, \ldots 30$, are \textit{parity-odd}, in the sense that they go to their inverse under $\sqrt{\Delta} \to - \sqrt{\Delta}$,
while all other letters are \textit{parity-even} under that transformation. Furthermore, only the letters $\{W_i\}_{i=1}^5\cup \{W_i\}_{i=16}^{20}$ can appear as first entries (see section~\ref{subsec: symbols} for information regarding the symbols). There is also a hypothetical second-entry condition that the integrals should obey that forbids some combinations of pairs of letters from appearing. We refer to \cite{Chicherin:2017dob} for more details.

\subsection{The canonical differential equations}

Using integration by parts identities (IBP), one can reduce the general integral
\eqref{eq:definition_of_F}, to a linear combination of a basis set of integrals $\vec{I}(v_i;\epsilon)$, called master integrals. The next step is then to find a way to compute those master integrals. We can accomplish this by using the method of differential equations~\cite{Kotikov:1990kg,Remiddi:1997ny,Gehrmann:1999as,Henn:2013pwa}, which works as follows. We first differentiate the set of master integrals $\vec{I}$ with respect to 
 the variables \eqref{eq: definition vi}. This can be done at the level of the integrals \eqref{eq:definition_of_F} by using appropriate derivatives in the external momenta, respecting the on-shell conditions. The derivatives obtained in this way can then also be expressed as a linear combination of the master integrals $\vec{I}$, which means that we obtain a set of first order linear differential equations
\begin{equation}
\frac{\partial \vec{I}(v_i;\epsilon)}{\partial v_j}\,=\,A_j(v_i;\epsilon)\vec{I}(v_i;\epsilon)\,,
\end{equation}
with five different matrices $\{A_j\}_{j=1}^{5}$ that depend in a non-trivial way on the $v_i$ as well as on $\epsilon=2-D/2$.  Now, if the set of master integrals is chosen to be of a d-log form, as is discussed in section~\ref{sec:dlog}, then the differential equations simplify significantly.  For such a basis,  after combining the five derivatives in a 1-form, we obtain the following canonical form of the differential equations \cite{Henn:2013pwa}
\begin{align}
\label{canonicalDEpentagon}
d \vec{I}(v_{i};\epsilon) \,=\, \epsilon \, d \tilde{A} (v_i) \, \vec{I}(v_{i};\epsilon) \,,
\end{align} 
with the matrix being independent of $\epsilon$.  We note that once \eqref{canonicalDEpentagon} and the value of $\vec{I}$ at some boundary point are known, then the problem of computing the master integrals $\vec{I}$ at any kinematic point in an $\eps$ expansion is solved \cite{Henn:2013pwa}.  The value of the integrals at the boundary point will be derived in section~\ref{sec:de}. We wish to emphasize here that the construction of the canonical basis is done at the \textit{integrand} level and as such does not require the a priori knowledge of the differential equation.

Finally, let us anticipate that one can write the $\tilde{A} (v_i)$ matrix in a nice way by using the algebraic functions $W_i$ of section~\ref{sec:kinematics} as
\begin{align}
\label{definition-Atilde}
\tilde{A}(v_i)\,=\,  \left[ \sum_{i=1}^{31} \tilde{a}_{i} \,  \log W_{i}(v_{i}) \right] \,,
\end{align}
where the $\tilde{a}_{i}$ are {\it constant} $73 \times 73$ matrices (with rational entries). We remark that $\tilde{A}$ is independent of seven of the letters, namely of the letters 8, 9, 10, 21, 22, 23 and 24. The corresponding $\tilde{a}_i$ matrices are zero.
%

\section{Construction of a basis of d-log integrals}
\label{sec:dlog}

In this section we describe how we obtained a d-log basis with constant leading singularities.
An algorithm for doing this is provided in \cite{WasserMSc}. 
Let us briefly summarize the method.
We start from a given propagator structure, in the present case that of part a) of Fig.~\ref{fig:boxfamily}.
Then, an ansatz for all possible numerator structures is made. The degree of the latter is constrained
by the requirement of the absence of double poles.
Computing all leading singularities of a general linear combination of such numerators, we obtain a complete solution of all d-log integrands for the corresponding propagator structure.

We perform the analysis in four dimensions, expressing the loop momenta in a basis built from the spinor helicity variables of the external momenta. 
Furthermore, we find it convenient to parametrize the kinematics as in eq. (3) of ref.~\cite{Bern:1993mq}, as the latter rationalizes the Gram determinant $\sqrt{\Delta}$ \eqref{eq: Gram Determinant} that can be built from four of the five external momenta.

The computation of the leading singularities can be combined nicely with the computation 
of cuts. In the case of the maximal cut of the full topology there are no integration variables left, so 
we obtain the leading singularities in this case without further calculations.
Computing the leading singularities on cuts has several advantages. First, it drastically reduces 
the amount of different leading singularities that have to be computed.
Second, we can split the calculation in several smaller parts that can be computed 
in parallel. Third, we can choose for each cut an optimized parametrization of the loop 
momenta and this way minimize the appearance of square roots in intermediate steps. 

In order to find a d-log solution in a given sector we proceed as follows: 
First, we compute the leading singularities on the maximal cut of that sector
in order to get all solutions projected on that sector. 
Then, for each solution we add a linear combination of integrals of the 
subsectors and fix their coefficients by computing the leading singularities on that subsector. In this way, we can iteratively construct a list of d-log integrals. 

As a check that the integrals obtained with this procedure are correct, we verified them all by computing the leading singularities for each solution individually without taking cuts.
Along this way we also checked that the solutions for the hexa-box family provided in \cite{Bern:2015ple} are d-log integrals with constant leading singularities.
For the verification we used a semi-numerical approach, setting \emph{all but one} of the external kinematical variables to numerical values, thus proving that the leading singularity does not depend on the one kinematical variable that was not replaced by a numerical value. Repeating this for all external variables ascertains 
that a given possible solution has constant leading singularities. This semi-numerical approach simplifies the calculation substantially.

We remark on a subtlety in this approach. As the above analysis is done in four dimensions,
it cannot detect certain Gram determinants that vanish in four dimensions. 
Therefore the latter represent an ambiguity.
While we expect a refined version of the leading singularity analysis to also fix
this ambiguity, here we chose a pragmatic solution.
We aimed for finding 'simple' solutions without the admixtures of Gram determinants (that necessarily involve many numerator terms,
and hence typically integrals of several topologies). Unwanted and complicated solutions of this type can in most cases be easily identified and removed.

In this way, we obtained  157 d-log integrals for the hexa-box family.
The integrals obtained are all expected to be pure functions of uniform transcendental weight \cite{ArkaniHamed:2010gh,Henn:2013pwa}.
We perform the following consistency check on this assertion. 
The number of d-log integrals 
(in our case 157) is much bigger than the number of master integrals (in our case 73).
We first choose a set of linearly independent d-log integrals as a basis
of master integrals.
Then, we express the remaining
integrals in terms of this basis, using the reduction obtained in Section~\ref{sec:kinematics} above.
If all integrals have uniform transcendental weight, then the basis coefficients must be numerical constants (for general Feynman integrals these coefficients would be functions of the external variables $s_{ij}=2p_i\cdot p_j$ and of $D$, the
parameter of dimensional regularization). Indeed, we explicitly found that all relations involved constants only.

\section{Determination of the boundary conditions}
\label{sec:de}

In this section, we will determine the boundary conditions of the differential equations \eqref{canonicalDEpentagon} for the hexa-box integrals, such that their complete solution becomes uniquely specified. 
 The method for computing the boundary conditions starts by picking a convenient reference point where the integrals are finite. Then, one integrates the differential equation  along a path joining the boundary point with kinematic points where letters of the alphabet vanish and where singularities can thus appear. By demanding the absence of spurious singularities, we obtain constraints on the values of the integrals at the reference point. This turns out to be sufficient to determine the boundary conditions (up to a trivial overall normalization).

\subsection{The origin point and spurious singularities}
The Euclidean region is given by the conditions 
\begin{equation}
v_1<v_3+v_4\,,\qquad v_2<v_4+v_5\,,\qquad v_i<0 \text{ for }i=3,4,5\,.
\end{equation}
One may verify that in this case the Feynman denominator of any integral of the integral family under consideration is positive definite. This implies that the solution of the differential equation \eqref{canonicalDEpentagon} is real within that region. The latter observation is useful, as the equations we will obtain are in general complex.

In order to provide an explicit solution to the differential equation, we need to determine the value of $\vec{I}(v_{i};\epsilon)$ at one point. A suitable candidate is the point $p_E$  in the Euclidean region corresponding to setting $v_1=-3$, $v_2=-3$, $v_3=v_4=v_5=-1$ and choosing the positive sign for the root, $\sqrt{\Delta}=3\sqrt{5}$. Let us denote the value of $\vec{I}(v_{i};\epsilon)$ at $p_E$ as $\vec{I}_E(\epsilon)$.

We can now impose conditions on the value of $\vec{I}_E(\epsilon)$ by demanding that the integrals stay finite when taking certain suitable limits. 
This is justified as follows: all integrals in the d-log basis are ultraviolet finite, by construction. We take $\epsilon<0$ in order to regulate the integrals in the infrared and consider limits in which some of the letters of the alphabet vanish. 
Taking such limits does not change the UV structure of the integrals, and so we require that the integrals remain finite in the limit, provided that $\epsilon<0$. In other words, we constrain the boundary condition 
$\vec{I}_E(\epsilon)$ by demanding that these \textit{spurious singularities} are absent.

\subsection{The paths to the spurious singularities }

Let us now explain how this is implemented in detail.
We begin by choosing an index $j\in\{1,\ldots, 25\}$ and considering the limit $W_j\equiv y\rightarrow 0$. Without loss of generality, let us take $j=11$ to illustrate the situation. We decompose the matrix $\tilde{A}$ as 
\begin{equation}
\tilde{A}=\tilde{A}_{\text{sing}} \log(y) + \text{non-singular for }y\rightarrow 0\,. 
\end{equation}
In the neighborhood of $y=0$, the solution of \eqref{canonicalDEpentagon} has the form
\begin{equation}
\label{eq: exponentiating the matrix at the boundary}
\vec{I}(v_{i};\epsilon)=\exp\left[\epsilon \log(y) \tilde{A}_{\text{sing}} \right] \vec{J}(\epsilon)+\mathcal{O}(y)\,,
\end{equation}
where $\vec{J}(\epsilon)$ is a constant boundary vector. Computing explicitly the matrix exponential in the above equation, we obtain many terms proportional to $y^{a\epsilon}$, where $a$ is an integer. Since we demand that the integrals are \textit{finite} at $y=0$ for \textit{negative values} of $\epsilon$, the coefficients of $y^{a\epsilon}$ in \eqref{eq: exponentiating the matrix at the boundary} have to vanish for $a>0$. This imposes conditions on the constant vector $\vec{J}(\epsilon)$, which we now have to translate to conditions on the value of the integral at the Euclidean point  $\vec{I}_E(\epsilon)$, see for example \cite{Henn:2013nsa}.

In order to do this translation, we need to consider a path $\gamma(x)$ that starts at $p_E$ and continues to a point $p$ in which $W_{11}$ vanishes. It is advantageous  to choose the parametrization of the path in such a way as to resolve the square root in $\sqrt{\Delta}$. Explicitly, the path reads
\begin{equation}
\label{eq: integration path}
v_1=-3\,,\quad v_2=-3\,,\quad v_3=-1\,,\quad v_4=-\frac{1}{4}(x^2-1)\,,\quad v_5=-1\,,\quad \sqrt{\Delta}=3x\,,
\end{equation}
where $x$ parametrizes the path. 
This path  reduces to $p_E$ for the beginning point $x=x_0=\sqrt{5}$ and leads to the vanishing of $W_{11}$ (and also some other letters) 
for the end point $x=x_1=\sqrt{13}$. 

\begin{figure}[htbp!]
             \begin{center}       
              \includegraphics[width=10cm]{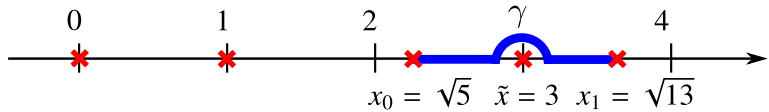}
              \caption{The integration path \eqref{eq: integration path}, going under the pole at $\tilde{x}=3$, is shown by the thick blue curve. Zeros of the letters \eqref{fullpentagonalphabet} are marked by red crosses. }
              \label{Fig:exampleIntegrationPath}
            \end{center}
\end{figure}

Along the path from $x_0$ to $x_1$, we have to go around the (spurious) singularity at $\tilde{x}=3$ where the letters 18, 
19, 27, 28 vanish. 
Since such a singularity can introduce a branch cut, we need to go around it by adding a small imaginary part. 
We can do in two ways, namely above or below the cut. Thus, in general, we obtain in principle two solutions for the value of the integrals in the vicinity of the end point $p$ and two corresponding path parameterizations. In practice, we do not need to worry about this and can take just one of the two, say the one going over the cut. We will then in general obtain complex equations for the unknown \textit{real} vector $\vec{I}_E(\epsilon)$, but we can simply decompose them into real and imaginary parts. 
We illustrate the path $\gamma$ in Fig.~\ref{Fig:exampleIntegrationPath}. 

We now expand the boundary values of the integral as $\vec{I}_E(\epsilon)=\sum_{n=0}^\infty  \vec{I}_{E}^{(n)}\epsilon^n$ and $\vec{J}(\epsilon)=\sum_{n=0}^\infty  \vec{J}^{(n)}\epsilon^n$. On the one hand, integrating the differential equation, we get the \textit{iterated integrals} expression
\begin{equation}
\label{eq: first expansion}
\vec{I}(v_{i};\epsilon)\,=\,\vec{I}_{E}^{(0)}+\epsilon\left(\int_{\gamma}d\tilde{A}\vec{I}_{E}^{(0)}+\vec{I}_{E}^{(1)}\right)+\epsilon^2\left(\int_{\gamma}d\tilde{A}\left(\int_{\gamma}d\tilde{A}\vec{I}_{E}^{(0)}+\vec{I}_{E}^{(1)}\right)+\vec{I}_{E}^{(2)}\right)+\cdots\,, 
\end{equation}
while on the other hand we get from  \eqref{eq: exponentiating the matrix at the boundary} the expansion \begin{equation}
\begin{split}
\label{eq: second expansion}
\vec{I}(v_{i};\epsilon)\,=\,&\vec{J}^{(0)}+\epsilon  \left(\tilde{A}_{\text{sing}} \vec{J}^{(0)} \log (y)+\vec{J}^{(1)}\right)\\&+\epsilon ^2 \left(\frac{1}{2} \tilde{A}_{\text{sing}}^2 \vec{J}^{(0)} \log^2(y)+\tilde{A}_{\text{sing}} \vec{J}^{(1)} \log (y)+\vec{J}^{(2)}\right)+\mathcal{O}(\epsilon^3)+\mathcal{O}(y)\,.
\end{split}
\end{equation}
In the above, we have to first impose on $\vec{J}(\epsilon)$ the vanishing of the terms proportional to $y^{a\epsilon}$ with $a>0$ in \eqref{eq: exponentiating the matrix at the boundary}. Furthermore, the parameter $y$ needs to be matched to the parametrization of the path as $y=x_1-x=\sqrt{13}-x$. The matching of \eqref{eq: first expansion} with \eqref{eq: second expansion} imposes conditions on the $\vec{I}_{E,n}$. 

We now need to evaluate explicitly the iterated integrals like $\int_{\gamma}d\tilde{A}\vec{I}_{E}^{(0)}$ in \eqref{eq: first expansion}. This task is performed explicitly in terms of Goncharov polylogarithms $G(a_1,\ldots, a_k;z)$ in three steps. First, we perform the iterated integrations along the path $\gamma(x)$ around the beginning point $x_0$ using the definitions of the Goncharov polylogarithms:
\begin{equation}
\label{eq: Goncharov definition}
G(\underbrace{0,\ldots,0}_{k};z)\,=\,\frac{1}{k!}\big(\log z\big) ^k\,,\qquad G(a_1,\ldots, a_k;z)\,=\,\int_{0}^z\frac{dt}{t-a_1}G(a_2,\ldots, a_k;t)\,.
\end{equation}
In a second step, 
approaching the end point $x_1$, we need to use the shuffle algebra for the Goncharov polylogarithms in order to make the terms containing $\log(y)=\log(x_1-x)$ explicit so that we can match \eqref{eq: first expansion} to \eqref{eq: second expansion}. This means that we obtain an explicit expression for the integrals like $\int_{\gamma}d\tilde{A}\vec{I}_{E}^{(0)}$ in the vicinity of $y=0$ that is of the type $\sum_m c_m \log(y)^m$ where the coefficients $c_m$ are $y$-independent and explicitly given in terms of the values of the Goncharov polylogarithms that are finite for $y=0$. The specific value of these constants depends in principle on the path chosen to avoid the spurious singularity at $\tilde{x}$. It suffices for our purposes to choose one of the two.  

We can now perform the matching \eqref{eq: first expansion} to \eqref{eq: second expansion} and obtain \textit{analytic} conditions on the $\vec{I}_{E}^{(n)}$ vectors that contain many different Goncharov polylogarithms. Repeating the same procedure that we did for the path $\gamma$ for many other paths going to other spurious singularities, we obtain many constraints on the boundary conditions.

Finally, we used one more constraint, which comes from the analysis of the leading singularities. One may classify the integral basis according to parity. Then, the parity odd integrals are expressed in terms of certain $F$'s of eq. (\ref{eq:definition_of_F}), and normalized by a factor proportional to $\sqrt{\Delta}$ to make them pure integrals. Since $\Delta \to  0$ is not a physical singularity of the Feynman integrals, the odd pure integrals have to vanish at $\Delta \to 0$. Similarly to the previous analysis we consider a path $\gamma(x)$, which rationalizes $\sqrt{\Delta}$,  joining the Euclidean point $p_E$ where $\Delta(p_E) = 3 \sqrt{5}$ and a singular point where $\Delta = 0$, and we integrate the differential equation along this path in terms of  Goncharov polylogarithms. The analytic conditions on the vector $\vec I_E^{(n)}$ come from vanishing of $\vec{I}(v_i;\epsilon)$ at the boundary point. More specifically, we use the shuffle algebra for the Goncharov polylogarithms to extract logarithmic singularities and we demand vanishing of the coefficients in front of all powers of the logarithms. Taking the union of all the constraints discussed above we find that they are sufficient to fix the boundary conditions analytically. We note that performing the matching at weight $L$ imposes additional conditions needed to fix the coefficients at weight $L-1$. Thus, we need to go to weight 5 for some of the paths, in order to obtain enough conditions to fix all the coefficients.

In fact, we obtain an overdetermined system of equations and solving it requires using many identities for the Goncharov polylogarithms. Thus, we choose to solve the equations numerically, which leads us to the third step, namely the numerical evaluation of those Goncharov polylogarithms that are finite at the end point $x_1$ of the path. In order do that, we use the \textsf{GiNaC} implementation of \cite{Vollinga:2004sn}. While in principle we can solve the equations to arbitrary numerical precision using \textsf{GiNaC}, to limit computing time, we chose 100 digits precision.

The consistency conditions from matching \eqref{eq: first expansion} to \eqref{eq: second expansion} for all the possible paths $\gamma$ going from $p_E$ to points at which some even letters $W_j$ vanish is enough to fix all the unknown coefficients in $\vec{I}_{E}^{(n)}$, up to an overall normalization condition. The latter reflects the fact that eq. \eqref{canonicalDEpentagon} is homogeneous in $\vec{I}(v_{i};\epsilon)$. To fix the normalization, it is 
sufficient to compute one of the trivial integrals in the d-log basis $\vec{I}(v_i;\epsilon)$ analytically. Factoring out 
the overall divergence and common factors from dimensional regularisation, the first component of  
$\vec{I}(v_{i};\epsilon)$ is defined and expressed as follows: 

\begin{align}
\label{eq: normalization of the integrals}
\vec{I}_1(v_{i};\epsilon)\,=\,&\epsilon^4 e^{2\epsilon \gamma_E}(-v_{5})F_{1,1,0,0,1,1,1,0,0,0,0}\nonumber\\
=&-(-v_{5})^{-2\epsilon}e^{2\epsilon \gamma_E}\frac{\Gamma(1-\epsilon)^3\Gamma(1+2\epsilon)}{4\Gamma(1-3\epsilon)}\\
=& (-v_5)^{-2 \epsilon}\left(-\frac{1}{4}+\frac{\pi ^2 \epsilon ^2}{24}+\frac{8 \zeta_{3} \epsilon ^3}{3}+\frac{19 \pi ^4 \epsilon ^4}{480}+O\left(\epsilon ^5\right)\right)\,.\nonumber
\end{align}
The result \eqref{eq: normalization of the integrals} provides the normalization fixing all the remaining coefficients.  In particular, we have for the first vector
\begin{equation}
\label{eq: result for IE0}
\begin{split}
\vec{I}_{E}^{(0)}\,=\,&\Big(-\frac{1}{4},0,0,\frac{1}{2},0,0,0,0,0,0,0,0,0,0,0,0,0,0,0,0,0,0,0,\frac{1}{4},0,0,\frac{1}{2},0,0,\frac{1}{4},\\&0,0,\frac{1}{2},0,0,0,0,0,0,0,0,0,0,0,0,0,0,-\frac{1}{4},0,\frac{1}{4},0,0,0,0,0,\frac{1}{8},-\frac{5}{4},-\frac{5}{4},\\&0,-\frac{1}{4},0,0,0,\frac{1}{4},\frac{1}{4},0,\frac{1}{4},0,0,0,-\frac{1}{4},-\frac{1}{4},-\frac{1}{4}\Big)\,.
\end{split}
\end{equation}

Furthermore, for illustration we show explicitly the complete solution for $\vec{I}(v_{i};\epsilon)$ up to linear order in $\epsilon$,
\begin{align}
 & \vec{I}(v_{i};\epsilon)\,=\, \vec{I}_{E}^{(0)} + \frac{\epsilon}{2} \Bigg(\log (-v_5),0,0,\log \left(\frac{1}{v_5^2}\right),0,0,\log \left(\frac{v_4}{v_2}\right),\log \left(\frac{v_2}{v_4}\right),0,0,\log \left(\frac{v_3^2}{v_1^2}\right),
\nonumber\\&
\log \left(\frac{v_3^2}{v_1^2}\right),0,0,0,0,0,0,2 \log \left(\frac{v_2}{v_2-v_4-v_5}\right),2 \log \left(\frac{v_2-v_4-v_5}{v_2}\right),
\nonumber\\&
2 \log \left(\frac{v_2}{v_2-v_4-v_5}\right),0,0,-\log (-v_1+v_3+v_4),\log \left(\frac{v_4}{v_1}\right),0,-2 \log (-v_1+v_3+v_4),
\nonumber\\&
0,0,-\log (-v_2+v_4+v_5),0,0,-2 \log (-v_2+v_4+v_5),0,\log \left(\frac{v_2}{v_4}\right),0,0,0,
\nonumber\\&
2 \log \left(\frac{v_1-v_3-v_4}{v_1}\right),0,0,0,0,0,\log \left(\frac{v_5^2}{v_2^2}\right),\log \left(\frac{v_5^2}{v_2^2}\right),0,\log (-v_3),\log \left(\frac{v_4}{v_1}\right),\log \left(\frac{-1}{v_4}\right),
\nonumber\\&
0,0,\log \left(\frac{v_1}{v_1-v_3-v_4}\right),\log \left(\frac{v_1-v_3-v_4}{v_1}\right),0,\log \left(\frac{v_3}{v_1}\right)+\frac{1}{2} \log \left(\frac{-v_2+v_4+v_5}{v_4 v_5}\right),
\nonumber\\&
\log \left(-\frac{1}{v_2^4 v_4}\right)+5 \log (v_5 (v_2-v_4-v_5)),\log \left(-\frac{1}{v_2^4 v_4}\right)+5 \log (v_5 (v_2-v_4-v_5)),
\nonumber\\&
0,\log (-v_4),\log \left(\frac{v_2}{v_2-v_4-v_5}\right),0,0,\log \left(\frac{v_2}{v_4 (-v_2+v_4+v_5)}\right),\log \left(-\frac{1}{v_4}\right),0,
\nonumber\\&
\log \left(-\frac{1}{v_4}\right),0,0,\log \left(\frac{v_2}{v_2-v_4-v_5}\right),\log \left(-\frac{v_3 v_4 v_5}{v_1 v_2}\right),
\nonumber\\&
\log \left(\frac{(v_1-v_3-v_4) v_4 (-v_2+v_4+v_5)}{v_1 v_2}\right),\log \left(-\frac{v_3 v_4 v_5}{v_1 v_2}\right)\Bigg)+\mathcal{O}(\epsilon^2)\,.
\end{align}
Inserting the values of $v_i$ for the Euclidean point $p_E$ in the above, one obtains our analytic expression for $\vec{I}_{E}^{(1)}$, which is proportional to $\log(3)$. As was already mentioned, the other boundary vectors $\vec{I}_{E}^{(n)}$ are fully determined by a system of equations involving Goncharov polylogarithms. The numerical solution to the latter is provided in an auxiliary file.
Numerical expressions for the boundary values up to weight 4 are listed in Tables~\ref{tab:boundary1} and \ref{tab:boundary2}.

Having fixed the boundary values up to weight 4 one can easily find an analytic solution of the differential equation (\ref{canonicalDEpentagon}) up to the same order in $\epsilon$-expansion.
Given a point in the Euclidean region of the kinematical space one connects it with the point $p_E$ by a path and integrates the differential equation along the path according to (\ref{eq: first expansion}). Choosing a piecewise linear path one can rationalize the integration kernels $d \tilde A$ and reduce all integrations to Goncharov polylogarithms \eqref{eq: Goncharov definition}. For more details on this procedure see e.g. \cite{Henn:2014lfa,Chicherin:2018ubl}.

\begin{table}
\centering
\begin{tabular}{ |c||c|c|c|c|c| } 
\hline
  & $I_{E}^{(0)}$ & $I_{E}^{(1)}$ & $I_{E}^{(2)}$ & $I_{E}^{(3)}$ & $I_{E}^{(4)}$ \\
 \hline
 \hline
$I_{1,\,E}$ &  $-{1\over4}$ &  0 &  0.4112335167 &  3.205485075 &  3.855776520  \\
\hline
$I_{2,\,E}$ &  0 &  0 &  0.4166359432 &  -1.078258215 &  1.041501311  \\
\hline
$I_{3,\,E}$ &  0 &  0 &  3.081680434 &  1.549694469 &  -0.3062825695  \\
\hline
$I_{4,\,E}$ &  ${1\over2}$ &  0 &  0.01080485305 &  -4.922299078 &  -4.628174866  \\
\hline
$I_{5,\,E}$ &  0 &  0 &  0.8224670334 &  0.6010284516 &  1.082323234  \\
\hline
$I_{6,\,E}$ &  0 &  0 &  0 &  -5.250469856 &  -17.31069279  \\
\hline
$I_{7,\,E}$ &  0 &  $-{1\over2}\log3$ &  -1.853382310 &  -2.567055582 &  2.373866827  \\
\hline
$I_{8,\,E}$ &  0 &  ${1\over2}\log3$ &  1.853382310 &  10.40068933 &  29.58205178  \\
\hline
$I_{9,\,E}$ &  0 &  0 &  1.228558667 &  1.496646401 &  1.938467722  \\
\hline
$I_{10,\,E}$ &  0 &  0 &  0.8116621804 &  2.610250132 &  0.9394466308  \\
\hline
$I_{11,\,E}$ &  0 &  $-\log3$ &  0.9771515547 &  4.821189178 &  9.345783210  \\
\hline
$I_{12,\,E}$ &  0 &  $-\log3$ &  5.287390655 &  6.658402302 &  -17.46337835  \\
\hline
$I_{13,\,E}$ &  0 &  0 &  0 &  -2.824257526 &  -4.481865549  \\
\hline
$I_{14,\,E}$ &  0 &  0 &  0 &  3.627039935 &  26.71708676  \\
\hline
$I_{15,\,E}$ &  0 &  0 &  0 &  -1.615301431 &  3.183030364  \\
\hline
$I_{16,\,E}$ &  0 &  0 &  -1.228558667 &  -1.496646401 &  -1.938467722  \\
\hline
$I_{17,\,E}$ &  0 &  0 &  0 &  -5.250469856 &  -17.31069279  \\
\hline
$I_{18,\,E}$ &  0 &  0 &  3.081680434 &  1.549694469 &  -0.3062825695  \\
\hline
$I_{19,\,E}$ &  0 &  $\log3$ &  -2.205710222 &  -5.108707833 &  17.15709578  \\
\hline
$I_{20,\,E}$ &  0 &  $-\log3$ &  0.9771515547 &  4.821189178 &  9.345783210  \\
\hline
$I_{21,\,E}$ &  0 &  $\log3$ &  -0.9771515547 &  -5.020211775 &  -5.172302363  \\
\hline
$I_{22,\,E}$ &  0 &  0 &  0.4166359432 &  -1.078258215 &  1.041501311  \\
\hline
$I_{23,\,E}$ &  0 &  0 &  0.3950262371 &  2.191861946 &  1.613550890  \\
\hline
$I_{24,\,E}$ &  ${1\over4}$ &  0 &  -0.4112335167 &  -3.205485075 &  -3.855776520  \\
\hline
$I_{25,\,E}$ &  0 &  $-{1\over2}\log3$ &  -1.853382310 &  -10.40068933 &  -29.58205178  \\
\hline
$I_{26,\,E}$ &  0 &  0 &  0 &  7.833633750 &  31.95591861  \\
\hline
$I_{27,\,E}$ &  ${1\over2}$ &  0 &  -2.654239637 &  -0.2598767588 &  -1.279639879  \\
\hline
$I_{28,\,E}$ &  0 &  0 &  0 &  1.615301431 &  -3.183030364  \\
\hline
$I_{29,\,E}$ &  0 &  0 &  0 &  -5.242341366 &  -23.53405639  \\
\hline
$I_{30,\,E}$ & ${1\over4}$  &  0 &  -0.4112335167 &  -3.205485075 &  -3.855776520  \\
\hline
$I_{31,\,E}$ &  0 &  0 &  0.4166359432 &  -1.078258215 &  1.041501311  \\
\hline
$I_{32,\,E}$ &  0 &  0 &  3.081680434 &  1.549694469 &  -0.3062825695  \\
\hline
$I_{33,\,E}$ &  ${1\over2}$ &  0 &  0.4274407963 &  1.289817710 &  -1.585922449  \\
\hline
$I_{34,\,E}$ &  0 &  0 &  0 &  -5.250469856 &  -17.31069279  \\
\hline
$I_{35,\,E}$ &  0 &  ${1\over2}\log3$ &  1.853382310 &  10.40068933 &  29.58205178  \\
\hline
$I_{36,\,E}$ &  0 &  0 &  0 &  -7.833633750 &  -31.95591861  \\
\hline
$I_{37,\,E}$ &  0 &  0 &  -1.228558667 &  -1.496646401 &  -1.938467722  \\
\hline
\end{tabular}
\caption{Numerical expressions for the boundary values  (integrals from 1 to 37).}
\label{tab:boundary1}
\end{table}

\begin{table}
\centering
\begin{tabular}{ |c||c|c|c|c|c| } 
\hline
  & $I_{E}^{(0)}$ & $I_{E}^{(1)}$ & $I_{E}^{(2)}$ & $I_{E}^{(3)}$ & $I_{E}^{(4)}$ \\
 \hline
\hline
$I_{38,\,E}$ &  0 &  0 &  -1.623584904 &  -3.688508347 &  -3.552018612  \\
\hline
$I_{39,\,E}$ &  0 &  $- \log3$ &  4.058831988 &  5.161755901 &  -19.40184607  \\
\hline
$I_{40,\,E}$ &  0 &  0 &  0 &  1.209127746 &  28.44134671  \\
\hline
$I_{41,\,E}$ &  0 &  0 &  0 &  -2.824257526 &  -4.481865549  \\
\hline
$I_{42,\,E}$ &  0 &  0 &  1.228558667 &  1.496646401 &  1.938467722  \\
\hline
$I_{43,\,E}$ &  0 &  0 &  0 &  -5.250469856 &  -17.31069279  \\
\hline
$I_{44,\,E}$ &  0 &  0 &  3.081680434 &  1.549694469 &  -0.3062825695  \\
\hline
$I_{45,\,E}$ &  0 &  $-\log3$ &  4.058831988 &  5.161755901 &  -19.40184607  \\
\hline
$I_{46,\,E}$ &  0 &  $-\log3$ &  4.058831988 &  4.962733304 &  -15.22836522  \\
\hline
$I_{47,\,E}$ &  0 &  0 &  0.4166359432 &  -1.078258215 &  1.041501311  \\
\hline
$I_{48,\,E}$ &  $-{1\over4}$ &  0 &  0.4112335167 &  3.205485075 &  3.855776520  \\
\hline
$I_{49,\,E}$ &  0 &  $-{1\over2}\log3$ &  -1.853382310 &  -10.40068933 &  -29.58205178  \\
\hline
$I_{50,\,E}$ &  ${1\over4}$ &  0 &  -0.4112335167 &  -3.205485075 &  -3.855776520  \\
\hline
$I_{51,\,E}$ &  0 &  0 &  1.228558667 &  1.496646401 &  1.938467722  \\
\hline
$I_{52,\,E}$ &  0 &  0 &  1.228558667 &  1.496646401 &  1.938467722  \\
\hline
$I_{53,\,E}$ &  0 &  ${1\over2}\log3$ &  1.041720130 &  -0.5462076011 &  -16.82418060  \\
\hline
$I_{54,\,E}$ &  0 &  $-{1\over2}\log3$ &  -1.041720130 &  2.280651020 &  27.50424540  \\
\hline
$I_{55,\,E}$ &  0 &  0 &  -6.297341812 &  -9.822049435 &  -3.068430467  \\
\hline
$I_{56,\,E}$ &  ${1\over8 }$ &  $- {1\over2 }\log3$ &  2.165967880 &  24.49213046 &  156.1420987  \\
\hline
$I_{57,\,E}$ &  $-{5\over4}$ &  $-2 \log3$ &  0.8474063649 &  24.27124243 &  153.1091184  \\
\hline
$I_{58,\,E}$ &  $-{5\over4}$ &   $-2 \log3$ &  13.71005188 &  29.93009110 &  -87.91862141  \\
\hline
$I_{59,\,E}$ &  0 &  0 &  0 &  -5.125173252 &  -62.08638519  \\
\hline
$I_{60,\,E}$ &  $-{1\over4}$ &  0 &  3.701101650 &  12.82194030 &  19.00830179  \\
\hline
$I_{61,\,E}$ &  0 &  ${1\over2}\log3$ &  1.458356073 &  -1.624465816 &  -15.78267929  \\
\hline
$I_{62,\,E}$ &  0 &  0 &  0 &  -1.734443419 &  -10.68006480  \\
\hline
$I_{63,\,E}$ &  0 &  0 &  0 &  1.054404157 &  -14.67727744  \\
\hline
$I_{64,\,E}$ & ${1\over4}$ &  ${1\over2}\log3$ &  -3.065473154 &  -31.98617740 &  -147.2653525  \\
\hline
$I_{65,\,E}$ &  ${1\over4}$ &  0 &  -2.878634617 &  -13.08813356 &  -23.26601096  \\
\hline
$I_{66,\,E}$ &  0 &  0 &  0 &  -8.275875993 &  -44.91759048  \\
\hline
$I_{67,\,E}$ &  ${1\over4}$ &  0 &  -2.878634617 &  -12.22091185 &  -17.92597856  \\
\hline
$I_{68,\,E}$ &  0 &  0 &  -2.457117334 &  5.372049170 &  35.28251373  \\
\hline
$I_{69,\,E}$ &  0 &  0 &  2.873753277 &  -7.317529094 &  -39.58104482  \\
\hline
$I_{70,\,E}$ &  0 &  ${1\over2}\log3$ &  -0.1868385372 &  -15.75813960 &  -146.0259443  \\
\hline
$I_{71,\,E}$ &  $-{1\over4}$ &  $-\log3$ &  6.051076583 &  57.14329049 &  290.2339876  \\
\hline
$I_{72,\,E}$ &  $-{1\over4}$ &  $-\log3$ &  -0.3802461731 &  13.16800937 &  75.27058162  \\
\hline
$I_{73,\,E}$ &  $-{1\over4}$ &  $-\log3$ &  -0.3802461731 &  13.16800937 &  75.27058162 \\
\hline
\end{tabular}
\caption{Numerical expressions for the boundary values (integrals from 38 to 73).}
\label{tab:boundary2}
\end{table}

\subsection{The symbol of the solution}
\label{subsec: symbols}

Thanks to the boundary vector \eqref{eq: result for IE0}, we can easily derive an explicit expression for the symbol of all of the 73 integrals. We refer to \cite{Duhr:2011zq} for a general introduction to symbols and to \cite{Chicherin:2017dob} for specific information concerning the integrable symbols relevant for this article. It follows directly from the differential equation \eqref{canonicalDEpentagon} and the definition \eqref{definition-Atilde}, that the symbol of the integrals we have computed are given by
\begin{equation}
\label{eq: symbols for all the integrals}
\begin{split}
\left[\vec{I}(v_{i};\epsilon)\right]&\,=\,\sum_{m=0}^\infty \epsilon^m\, [\vec{I}^{(m)}(v_{i};\epsilon)]\\
\text{ with }\,  [\vec{I}^{(m)}(v_{i};\epsilon)] &\,=\,\sum_{i_1,\ldots, i_m=1}^{31}\tilde{a}_{i_m}\cdot \tilde{a}_{i_{m-1}}\cdots \tilde{a}_{i_1}\cdot \vec{I}_{E}^{(0)} \left[W_{i_1},\ldots, W_{i_m}\right]\,.
\end{split}
\end{equation}
It should be noted that  the standard ordering in symbols is the opposite to that of the Goncharov polylogarithms in \eqref{eq: Goncharov definition}. For symbols $[W_{i_1},\ldots, W_{i_n}]$, derivatives act on the last entry.


\subsection{Checks on the solution}

Several independent checks were performed on the hexa-box integrals derived above. The full set of integrals can 
be compared to the purely numerical evaluation, obtained using 
sector-decomposition with the  \textsf{FIESTA} \cite{Smirnov:2015mct} code. The comparison is performed in the 
 Euclidean point $p_E$, and yields good agreement within the available numerical precision. However, the error 
margins  on the \textsf{FIESTA} results increase with increasing weight, and agreement  can be established for 
 $\vec{I}_{E}^{(3)}$ only to 1$\%$ and for  $\vec{I}_{E}^{(4)}$ only to 2$\%$.

The  hexa-box integral family contains subtopologies corresponding to 
planar and non-planar four-point functions with one 
off-shell leg~\cite{Gehrmann:2000zt, Gehrmann:2001ck, Gehrmann:2002zr} and to planar
 five-point functions~\cite{Gehrmann:2015bfy,Papadopoulos:2015jft,Gehrmann:2018yef}. Analytical expressions for 
 all these integrals were derived previously. Working again in the Euclidean point $p_E$, we performed a detailed
 numerical comparison for all previously available integrals (63 of the 73 integrals from the hexa-box family), using the routines 
 described in~\cite{Gehrmann:2001pz, Gehrmann:2001jv}  for the four-point functions and~\cite{Gehrmann:2018yef} 
 for the five-point functions, observing  full agreement of the results. It is worth noting that the 
 Euclidean five-particle kinematics translates for some of 
 the subsector integrals into (space-like) Minkowskian four-particle kinematics~\cite{Gehrmann:2002zr}, where the 
 integrals nevertheless remain real. 
 
Finally, in appendix~\ref{sec:mellinbarnes} we perform a direct check of the symbols of some of the components of $\vec{I}(v_{i};\epsilon)$ by deriving their Mellin-Barnes representation, which can then be used to bootstrap their symbol using the methods explained in \cite{Chicherin:2017dob}. We obtain a perfect agreement with the expression in \eqref{eq: symbols for all the integrals}.

\section{Conclusion and discussion}
\label{sec:conclusion}

In this paper, we computed an analytical expression for 
all massless non-planar two-loop five-point integrals belonging to the  hexa-box integral 
family. Our computation is based on the identification of a basis of integrals with constant 
leading singularities, which fulfil a system of differential equations in canonical form. Inspection of this 
system verifies a conjecture 
made in \cite{Chicherin:2017dob} about the function space governing these integrals.
By construction, the d-log form of the differential equation system is solved trivially in terms of iterated integrals. 

To uniquely determine the integrals from their differential equations requires knowledge on their boundary 
values in one specific kinematical point. Using physical insights on the singularity structure of the integrals, 
we infer boundary conditions from their behaviour in spurious kinematical points where the differential equations become 
singular, but the integrals themselves should remain regular. These boundary conditions are combined 
into a boundary value for all integrals in one specific point in the Euclidean region, from where 
the  integrals can be evaluated straightforwardly, for example in terms of Goncharov polylogarithms.

The integrals are real and single-valued in the Euclidean region. For practical applications in scattering 
amplitude calculations, their analytical continuation to the Minkowskian regions corresponding to 
all kinematical crossings is required. This can 
in principle be performed on the iterated integrals with an appropriate deformation of the integration contours. 
Aiming for an efficient numerical representation in all  regions, a more systematic approach is in order, analogous to
the work on the planar two-loop five-point integrals~\cite{Gehrmann:2018yef}. In there, the minimal basis of 
planar pentagon functions was identified from their required analyticity properties, expressed in entry conditions 
on their symbol. All these functions were written in terms of one-dimensional integrals containing simple logarithmic  
and polylogarithmic integrands, with boundary values determined separately in each Minkowskian region. 
A similar procedure should equally be feasible for the non-planar five point integrals from the hexa-box family. It
will be subject of future work, aiming for an efficient numerical representation for arbitrary physical kinematics.

The master integrals considered in this paper are relevant for two-to-three scattering processes in arbitrary theories with massless particles. 
Any integral of the hexa-box family can be expressed in terms of the master integrals computed here. In some cases, this may involve additional integral reduction identities beyond the ones used here for deriving the differential equations. 
There are by now several approaches~\cite{Boehm:2018fpv,Chawdhry:2018awn} for finding such integral reductions.
On the other hand, in the case of five-particle scattering in $\cN = 4$ super Yang-Mills theory, no further integral reductions are necessary. This is due to the fact that all hexa-box integrals appearing there are directly part of our integral basis.
 Therefore, with the work presented here, all integrals of one of the 
two non-planar integral families contributing to the amplitude \cite{Bern:2015ple} are known. The second non-planar integral family is beyond the scope of the present paper.

\bigskip

\noindent {\bf Note added:}
After completion of this paper, important related progress was made by the analytic reconstruction of two-loop five-gluon QCD amplitudes \cite{Badger:2018enw,Abreu:2018zmy} at leading color level in terms of master integrals, and the analytic determination of the full-color five-point two-loop amplitude in $\cN=4$ SYM theory \cite{Abreu:2018aqd,Chicherin:2018yne} at symbol level. Moreover, the calculation of the integrals \cite{Abreu:2018aqd,Chicherin:2018old} from the double pentagon family (Fig.~\ref{fig:boxfamily}b) completed the full set of massless two-loop five-point master integrals.

\section*{Acknowledgements}
This work was supported in part by the Swiss National Science Foundation (SNF) 
under contract  200020-175595, by the PRISMA Cluster of Excellence at Mainz University
and by the European Research Council (ERC)  under grants
 {\it MC@NNLO} (340983) and {\it Novel structures in scattering amplitudes} (725110).

\appendix

\section{Comparison with Mellin-Barnes calculation}
\label{sec:mellinbarnes}

In this appendix, we compute the symbols of several integrals using the Mellin-Barnes technique, which provides a useful check of the results of the main text. 
The integrals we discuss are also of direct importance for amplitudes in $\cN=4$ super Yang-Mills theory.

\label{subsec: MB representation explanation}

We are interested in checking the symbols \eqref{eq: symbols for all the integrals}. For this, we shall compute the symbols  of a few integrals using the Mellin-Barnes bootstrap technique of \cite{Chicherin:2017dob}.  The integrals that we shall consider are the four members of the hexa-box integral family that can be found in  \cite{Bern:2015ple}, see  Fig.~\ref{Fig:non-planar integrals}. In the notation of \cite{Bern:2015ple}, these are the integral (i), the integrals (h) with numerators $N^{(h)}_1 = \vev{15} [45] [12] \vev{2|(q-p_1)q|4}$ and $N^{(h)}_3 = s_{12} \vev{14}[15] \bra{5}q|4]$, and finally the integral (c) with the  numerator $N^{(c)} = \vev{15}[54]\vev{43}[1|q(q+p_{4}+p_5)|3] (q+p_4)^2$. The symbol of the integral (i) was computed in \cite{Chicherin:2017dob} up to and including the finite part in the $\ep$-expansion by means of the Mellin-Barnes technique. Here we outline how to obtain the symbols of integrals (h) and (c) by this method as well. In order to be more specific, let us from now on concentrate on the integral of type (h) with the numerator $N^{(h)}_1$, which we dub $I^{(h_1)}$. The steps that we perform can be done, with minimal modifications for the other (h) integral as well as for the integral (c).

\begin{figure}
\begin{center}
\includegraphics[width = 7cm]{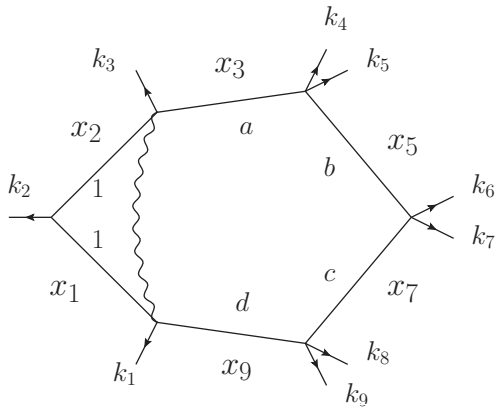}
\end{center}
\caption{One-loop hexagon integral with chiral numerator $\vev{1|x_{10}x_{02}|3}$ in region-momenta notations, $\ket{\lambda_i}[\tilde{\lambda}_i| = k_i =x_{i-1} - x_i$. The loop-integration $x_0$ is $D$-dimensional, $D=4-2\ep$, and the chiral numerator is four-dimensional. Pairs of on-shell momenta are used to represent an off-shell momentum.}
\label{fighexagontw}
\end{figure}

We start by deriving a neat Feynman representation for $I^{(h_1)}$, 
which will then allow us to obtain its Mellin-Barnes representation. This is done by using the fact that $I^{(h_1)}$ contains a box sub-integral (see Fig.~\ref{Fig:non-planar integrals}) which can be rewritten as a two-fold integral of a propagator raised to the power $2+\ep$ as:
\begin{align}
\pi^{\frac{D}{2}} \frac{\Gamma(-\ep)^2 \Gamma(2+\ep)}{\Gamma(-2\ep)} 
\int_0^1 d\tau \int_0^1 d \sigma \frac{1}{((\ell+\tau p_4 + \sigma p_5)^2)^{(2+\ep)}}\,.
\end{align}
In this way, we reduce the non-planar two-loop integral to a one-loop integral with non-integer indices of propagators. This integral is a special case of the following one-loop hexagon with a 'magic numerator', written here in the region-momenta notations and depicted in Fig.~\ref{fighexagontw}:
\begin{align}
\label{eq: first representation of Jhex}
J_{\rm hex}=\int d^D x_{0} \frac{\vev{\lambda_1|x_{10}x_{02}|\lambda_3}}{x_{10}^2 x_{20}^2 x_{30}^{2a} x_{50}^{2b} x_{70}^{2c} x_{90}^{2c}}\,,
\end{align}
where ones needs to relate the momenta $p_i$ to the $k_j$ of Fig.~\ref{fighexagontw} appropriately. 
By using momentum-twistors, similarly to \cite{ArkaniHamed:2010gh} though generalizing to $D$ dimensions, and by representing the numerator of \eqref{eq: first representation of Jhex} as a suitable derivative, one can derive a neat Feynman representation for the one-loop hexagon $J_{\rm hex}$:
\begin{align}
\label{mhexgn} 
J_{\rm hex}\,=\,& \frac{\pi^{\frac{D}{2}}\Gamma(a+b+c+d+\ep)}{\Gamma(a)\Gamma(b)\Gamma(c)\Gamma(d)}\int[d \beta] \beta_3^{a-1} \beta_5^{b-1} \beta_7^{c-1} \beta_9^{d-1} \notag\\&\times \biggl( 
\frac{\beta_5 \vev{\lambda_1|x_{15} x_{53}|\lambda_3}  + \beta_7 \vev{\lambda_1|x_{17} x_{73}|\lambda_3}}{[\sum_{k<l} \beta_k \beta_l x_{k l}^2 ]^{a+b+c+d+\ep}} \\
& +\frac{a+b+c+d-3+2\ep}{a+b+c+d-1+\ep}\frac{\vev{\lambda_1\lambda_3}}{[\sum_{k<l} \beta_k \beta_l x_{k l}^2 ]^{-1+a+b+c+d+\ep}}\biggr) \,.\notag
\end{align}
In the above, we have defined $[d\beta] = \delta(-1+\sum_k \beta_k) \prod_k d\beta_k$ with the indices $k,l$ taking the values $1,2,3,5,7,9$. Inserting now into \eqref{mhexgn} the appropriate parameters of $I^{(h_1)}$, namely $a=1$, $b=2+\epsilon$, $c=d=0$, we  obtain
\begin{multline}
\label{eq: Feynman parametrization Ih1}
I^{(h_1)} = \pi^D \vev{15}[35][12] \frac{\Gamma^2(-\ep)\Gamma(3+2\ep)}{\Gamma(-2\ep)}
\biggl[ 
\vev{45}[53]\vev{32} J_{1}^{(h_1)}\\+ v_4\vev{42} J_{2}^{(h_1)} 
+ \frac{3\ep}{2+2\ep} \vev{42} J_{3}^{(h_1)} \biggr]
\end{multline}
where we have defined the following three integrals over Feynman parameters
\begin{align}
\label{eq: Ji definitions}
J_{1}^{(h_1)} = \int d\Omega \, \frac{\sigma\alpha_4^{2+\ep}}{\big(F^{(h_1)}\big)^{3+2\ep}}\,, \quad 
J_{2}^{(h_1)} = \int d\Omega\,  \frac{\sigma\bar\tau\alpha_4^{2+\ep}}{\big(F^{(h_1)}\big)^{3+2\ep}} \,,
\quad 
J_{3}^{(h_1)} = \int d\Omega\, \frac{\alpha_4^{1+\ep}}{\big(F^{(h_1)}\big)^{2+2\ep}}\,.
\end{align} 
In \eqref{eq: Ji definitions}, we have used the shorthand $ d\Omega \equiv \delta\left(-1+\sum_{i=1}^4 \alpha_i\right)\, d\tau\, d \sigma\prod_{i=1}^4 d\alpha_i$ and the integration is performed over the domain $0 < \tau < 1$, $0 <\sigma < 1$ and $0 < \alpha_i < + \infty$. Note that $\bar\tau \equiv 1-\tau$ and $\bar\sigma = 1-\sigma$. Furthermore, the  $F$-polynomial of \eqref{eq: Ji definitions} is given explicitly as follows (note that $s_{ij}=2p_i\cdot p_j$)
\begin{equation}
\begin{split}
F^{(h_1)} =& \alpha_1 \alpha_3 s_{12} + \alpha_1 \alpha_4 \tau \sigma s_{45} + \alpha_2 \alpha_4 (\tau \sigma s_{45} + \tau s_{14} + \sigma s_{15}) \\&+ \alpha_3 \alpha_4 (\bar\tau\bar\sigma s_{45} +\bar\tau s_{34} + \bar\sigma s_{35})\,.
\end{split}
\end{equation}  
Using the Feynman representation \eqref{eq: Feynman parametrization Ih1}, we can obtain a Mellin-Barnes representation for $I^{(h_1)}$.
All we need to do is to use the basic Mellin-Barnes integral formula, 
\begin{align}\label{MBone}
\frac{1}{(X+Y)^a} = \frac{1}{\Gamma(a)} \int_{c-i \infty}^{c+i\infty} \frac{dz}{2\pi i} \Gamma(-z) \Gamma(a+z) X^z Y^{-a-z} \,,
\end{align}
where the $z$-integration goes along the vertical axis with real part $c\in(-a,0)$, and to then carry out the Feynman parameter integrals. 
In doing so we consider the $F$-polynomial $F^{(h_1)}$, not directly as a function of the $v_i$ of \eqref{eq: definition vi}, but rather equivalently as a function of the following five independent Mandelstam invariants, 
\begin{align} 
\label{invar}
s_{14}=v_2 - v_4 - v_5,\quad s_{15}=v_5,\quad s_{34}=v_3,\quad s_{45}=v_4,\quad s_{35}=v_1 - v_3 - v_4\,.
\end{align}
The explicit Mellin-Barnes representation for the $J_1^{(h_1)}$ piece of the integral, see \eqref{eq: Ji definitions}, reads
\begin{align}
\label{eq: MB integral for h1}
J_1^{(h_1)}\,=\,&\int\frac{\prod_{s=1}^9d z_s}{(2\pi i)^9}\frac{(-s_{14})^{z_4} (-s_{15})^{z_6} (-s_{34})^{z_1 + z_5}
    (-s_{35})^{-3 - 2 \epsilon - \sum_{s=1,3,4,5,6,9}z_s}(- s_{45})^{z_3 + z_9}
    }{\Gamma(-3 \epsilon) \Gamma\Big(-2 \epsilon - \sum_{s=1}^5z_s\Big) \Gamma\Big(-1 - 2 \epsilon - \sum_{s=1,2,3,6,9}z_s\Big)}\nonumber\\
    &\times 
     \left[\prod_{s=1}^9\Gamma(-z_s)\right]\Gamma\Big(-\epsilon -\sum_{s=1}^3z_s\Big)  \Gamma\Big(1 + \sum_{s=1,2,3,7}z_s\Big) \Gamma\Big(-2 - 2 \epsilon - \sum_{s=4,6,7,8}z_s\Big) \nonumber\\
    &\times \Gamma\Big(-2 - 2 \epsilon - \sum_{s=1}^8z_s\Big)\Gamma\Big(1 + \sum_{s=4,6,8}z_s\Big) \Gamma\Big(
    1 + \sum_{s=4,7,8}z_s\Big) \Gamma\Big(
    2 + \sum_{s=6}^8z_s\Big) \\
    &\times \Gamma\Big(-2 - 2 \epsilon - \sum_{s\neq 5}z_s\Big)  \Gamma\Big(3 + 2 \epsilon + \sum_{s=1}^9z_s\Big)\,,\nonumber
\end{align}
with similar expressions for the remaining $J_i^{(h_1)}$. Thus, we obtain a Mellin-Barnes representation for $I^{(h_1)}$. Since the five variables $s_{ij}$ of \eqref{invar} are negative in the Euclidean region and all terms of the polynomial $F^{(h_1)}$ are explicitly negative, the nine-fold Mellin-Barnes integrals are well defined in the Euclidean region. 
Now, expressing \eqref{eq: MB integral for h1} directly in terms of known functions would be very difficult. Fortunately, the Mellin-Barnes integrals simplify significantly when various kinematical limits are taken and we can exploit this in order to compute the symbol of the integral $\text{SB}[I^{(h_1)}]$. 

We compute the symbol of the integral by bootstrapping a suitable ansatz. The $\ep$-expansions of integral $I^{(h_1)}$ is of the form:
\begin{align} 
\label{intexp}
& I^{(h_1)} = \frac{1}{\ep^4} I^{(h_1)}_{0} + \frac{1}{\ep^3} I^{(h_1)}_{1} + \frac{1}{\ep^2} I^{(h_1)}_{2} 
+ \frac{1}{\ep^1} I^{(h_1)}_{3} + I^{(h_1)}_{4} + {\cal O}(\ep) \,,
\end{align}
where the $I^{(h_1)}_{k}$ are weight $k$-functions. We take an ansatz which is a linear combination of weight-$k$ integrable symbols whose seven allowed first entries correspond to the allowed unitarity cuts of the integrals $I^{(h_1)}$. Furthermore, we also impose the second entry condition conjectured in~\cite{Chicherin:2017dob}. The size of the ans\"atze, i.e. the number of even/odd symbols, is shown in Tab.~\ref{tabAns}. For example, at weight 4, we need to fix a priori $970+106=1076$ coefficients, in order to bootstrap the symbol of $I^{(h_1)}_{4}$. The symbols for the bootstrapping of the integral (c) are the same.

\begin{table}
\begin{center}
\begin{tabular}{l|c|c|c|c|c}
Weight & 0 & 1 & 2 & 3 & 4 \\ \hline
Size of ansatz & $1|0$ & $7|0$ & $36|1$ & $182|12$ & $970|106$ \\ \hline
\end{tabular}
\end{center}
\caption{Number of even$|$odd integrable symbols with seven allowed first entries and satisfying the second entry condition.} \label{tabAns}
\end{table} 

In order to fix all these coefficients,  we take various kinematical limits in which the Mandelstam invariants \eqref{invar} approach zero or infinity. The fact of taking such limits, simplifies the Mellin-Barnes integrals for the $J^{(h_1)}_i$, like \eqref{eq: MB integral for h1}, significantly by lowering their dimensionality, i.e.\ by reducing the number of contour integrals needed. We are interested in the limits for which the simplified Mellin-Barnes integrals can be evaluated explicitly by means of the Cauchy theorem. Furthermore, the same limits considerably simplify the 31-letter alphabet. We specialize to those limits leading to 2dHPL and HPL alphabets. Then, by considering the computed asymptotics of the Mellin-Barnes integrals and by comparing them to the symbol ansatz, we can fix the unknown coefficients in the ansatz. In this way, we obtain the symbol of the integral $I^{(h_1)}$ up to and including the finite part. The first few terms of it are explicitly
\begin{equation}
\label{eq: symbol for I^{(h_1)}}
[I^{(h_1)}]=
 \frac{1}{8}
 +\epsilon  \Big(-\frac{1}{2} \left[W_1\right]
 +\frac{\left[W_3\right]}{2}+\frac{\left[W_{19}\right]}{4}
 -\frac{\left[W_4\right]}{4}-\frac{\left[W_5\right]}{4}\Big)+\mathcal{O}(\epsilon^2)\,,
\end{equation}
but we stress that we computed all the terms up to and including the $\epsilon^4$ terms. 

To summarize, we obtain the symbol of $I^{(h_1)}$ by first deriving a Feynman representation by getting rid of the box sub-integral, then trading that Feynman representation for a Mellin-Barnes one which is very convenient for taking suitable kinematical limits for which the integral can be evaluated explicitly such that finally one obtains constraints for an inspired symbol ansatz.  The symbol \eqref{eq: symbol for I^{(h_1)}} can now be compared directly to the results we have obtained in the main text. Specifically, $I^{(h_1)}=-\big(\vec{I}(v_{i};\epsilon)\big)_{56}$ and we obtained the symbol of the right hand side in \eqref{eq: symbols for all the integrals}. We find that both sides are in complete agreement. 
 
Identical calculations for the symbols of the other integral of type (h) as well as of the integral (c) have also been done.
They are given in terms of the integrals as $I^{(h_3)}=\big(\vec{I}(v_{i};\epsilon)\big)_{57}$ and $I^{(c)}=\big(\vec{I}(v_{i};\epsilon)\big)_{71}$, and the symbol results agree with the computations of the main text.

 \bibliographystyle{JHEP}

\bibliography{bibfile_pentagon}

\end{document}